\documentclass[aps,pra,twocolumn,amsmath,amssymb,showpacs,floatfix,superscriptaddress,nofootinbib]{revtex4-2}

\usepackage{xcolor}
\usepackage[utf8x]{inputenc}
\usepackage[english]{babel}
\usepackage[T1]{fontenc}
\usepackage{lmodern}
\usepackage{amssymb}
\usepackage{amsthm}
\usepackage{amsmath}
\usepackage[pdftex]{graphicx}
\usepackage{epstopdf}
\usepackage{braket}
\usepackage[bottom]{footmisc}
\usepackage{bbold}
\usepackage{wasysym}
\usepackage{dcolumn}
\usepackage{bm}
\usepackage{color}
\usepackage{relsize,dsfont,mathrsfs,empheq,verbatim,upgreek}
\usepackage{etoolbox}
\usepackage[caption = false]{subfig}
\usepackage{mathtools}

\DeclarePairedDelimiter\norm{\lVert}{\rVert}%

\providecommand{\norm}[1]{\lVert#1\rVert}

\newcommand{\Tr}{\mathrm{Tr}}
\newcommand{\Lt}{\mathcal{L}}

\newcommand{\Tau}[2]{\boldsymbol{\tau}_{#1}^{#2}}
\newcommand{\Svec}[2]{\boldsymbol{s}_{#1}^{#2}}

\newcommand{\Dcal}[4]{\mathcal{D}\left(\Tau{#1}{#2}, \Svec{#3}{#4}\right)}
\newcommand{\Dfd}[4]{\dot{D}\left(\Tau{#1}{#2}, \Svec{#3}{#4}\right)}
\newcommand{\Df}[4]{{D}\left(\Tau{#1}{#2}, \Svec{#3}{#4}\right)}

\newcommand{\Dcalt}[2]{\mathcal{D}\left(\Tau{#1}{#2}\right)}
\newcommand{\Dfdt}[2]{\dot{D}\left(\Tau{#1}{#2}\right)}
\newcommand{\Dft}[2]{{D}\left(\Tau{#1}{#2}\right)}

\newcommand{\Dcals}[2]{\mathcal{D}\left(\Svec{#1}{#2}\right)}
\newcommand{\Dfds}[2]{\dot{D}\left(\Svec{#1}{#2}\right)}
\newcommand{\Dfs}[2]{{D}\left(\Svec{#1}{#2}\right)}

\newcommand{\Xcal}[5]{{#1}\!\left(\Tau{#2}{#3}, \Svec{#4}{#5}\right)}

\begin{document}
%%%%%%%%%%%%%%%%%%%%%%%%%%%%%%%%%%%%%%%%%%%%%

\title{Recursive perturbation approach to time-convolutionless master equations:\\
Explicit construction of generalized Lindblad generators for arbitrary open systems}
%%%%%%%%%%%%%%%%%%%%%%%%%%%%%%%%%%%%%%%%%%%%%

\author{Alessandra Colla}
\email{alessandra.colla@unimi.it}
\affiliation{Dipartimento di Fisica ``Aldo Pontremoli'', Universit\`a degli Studi di Milano, Via Celoria 16, I-20133 Milan, Italy}
\affiliation{INFN, Sezione di Milano, Via Celoria 16, I-20133 Milan, Italy}
\affiliation{Institute of Physics, University of Freiburg, 
Hermann-Herder-Stra{\ss}e 3, D-79104 Freiburg, Germany}
 
\author{Heinz-Peter Breuer}

\affiliation{Institute of Physics, University of Freiburg, 
Hermann-Herder-Stra{\ss}e 3, D-79104 Freiburg, Germany}

\affiliation{EUCOR Centre for Quantum Science and Quantum Computing,
University of Freiburg, Hermann-Herder-Stra{\ss}e 3, D-79104 Freiburg, Germany}

\author{Giulio Gasbarri}
\email{giulio.gasbarri@uni-siegen.de}
\affiliation{Naturwissenschaftlich-Technische Fakult\"{a}t, Universit\"{a}t Siegen, Siegen 57068, Germany}
\affiliation{F\'isica Te\`orica: Informaci\'o i Fen\`omens Qu\`antics, Department de F\'isica, Universitat Aut\`onoma de Barcelona, 08193 Bellaterra (Barcelona), Spain}
\begin{abstract}
We develop a recursive perturbative expansion for the time-convolutionless (TCL) generator of an open quantum system in a generalized Lindblad form. This formulation provides a systematic approach to derive the generator at arbitrary order while preserving a Lindblad-like structure, without imposing assumptions on the system or environment beyond an initially uncorrelated state. The generator is written, at all orders, in a canonical form, which also corresponds to the minimal dissipation condition, which uniquely specifies the decomposition of the generator into Hamiltonian and dissipative contributions.
To validate the method and show its effectiveness in addressing non-Markovian dynamics and strong-coupling effects, we compute the generator explicitly up to fourth order.
\end{abstract}

\maketitle
\section{Introduction}

The theory of open quantum systems provides a foundational framework for describing quantum dynamics in realistic environments, where system-environment interactions lead to energy and information exchange, often resulting in decoherence and irreversible behavior~\cite{Breuer2007, Vacchini2024}.

A widely used framework for describing such dynamics is the Lindblad master equation, which can be derived either phenomenologically or from a microscopic system-environment model. This formalism assumes Markovian dynamics, which holds under weak coupling and a clear separation of timescales between the system and the environment. However, in the presence of strong coupling or structured spectral densities, memory effects become significant, and the Lindblad equation no longer provides an accurate description.

Time-convolutionless (TCL) master equations provide a robust framework for modeling non-Markovian dynamics in open quantum systems by utilizing time-local generators that account for environmental back-action. However, their analytical derivation is generally limited to a few solvable models, typically involving Gaussian environments or linear system-bath interactions. Notable examples include the Jaynes-Cummings model~\cite{Smirne2010}, pure dephasing models~\cite{Luczka1990, Doll2008}, and Gaussian reservoirs~\cite{Diosi2014, Ferialdi2016}, as well as paradigmatic models such as quantum Brownian motion~\cite{Hu1992} and the Fano-Anderson model~\cite{Huang2022, Picatoste2024, Colla2024roles}.

In complex scenarios such as systems interacting with non-Gaussian reservoirs or governed by nonlinear couplings, the time-convolutionless (TCL) generator is typically derived through perturbative expansions. Traditional methods -- including projection operator techniques, stochastic methods~\cite{Shibata1977, Chaturvedi1979,strunz2004convolutionless}, and other analytical/numerical~\cite{deVega2017} strategies -- often grow intractable at higher orders due to the rapidly increasing complexity of nested time integrals and environment correlation functions.

To address this issue, a recursive perturbative expansion was developed in~\cite{Gasbarri2018}, systematically constructing the TCL generator using commutators and anticommutators of system operators and bath correlation functions. This method recursively yields higher-order terms from lower-order contributions, avoiding redundant calculations and enabling a diagrammatic representation of perturbative terms. By explicitly incorporating multi-time correlation functions, the approach handles non-Gaussian effects and nonlinear couplings, significantly improving the tractability of non-Markovian dynamics in open quantum systems. However, convergence of the perturbative series depends on system parameters, and singularities in the exact TCL generator can arise in specific regimes~\cite{vacchini2010exact}.

 In this work, we extend this framework by developing a recursive perturbative expansion of the TCL generator expressed in terms of left- and right-acting system operators. This representation clarifies the structural decomposition of the generator, enabling the identification of its canonical form, which consists of a coherent (Hamiltonian) component and a dissipative component that has the structure of a Lindblad dissipator but with time-dependent and potentially negative rates~\cite{Gorini1976a, Breuer2012, Hall2014}. The emergence of negative rates is a signature of non-CP-divisibility, indicating that the evolution cannot be represented as a sequence of completely positive propagators, a signature of potential non-Markovian dynamics~\cite{Rivas_2014, Breuer2016C}.

 While the decomposition into coherent and dissipative components is generally not unique~\cite{Breuer2007}, uniqueness can be enforced by requiring the Lindblad operators to be traceless -- a minimal dissipation condition that ensures the dissipator is minimal (as a superoperator) with respect to a Hilbert-Schmidt-averaged norm~\cite{Sorce2022}. Although this criterion is not strictly necessary, it yields a well-defined canonical decomposition, resolving ambiguities inherent to non-Markovian regimes.

The coherent part of the TCL master equation in this canonical form -- and, in particular, the effective system Hamiltonian generating it -- has received significant attention, as it describes the renormalization of the system's energy levels due to interaction with the environment.
While in the Markovian case this contribution corresponds to the well-known Lamb shift~\cite{Bethe1947,Lamb1947,Breuer2007}, in the non-Markovian and strong-coupling regime the renormalization is generally time-dependent and can be much more pronounced. Indeed, recent experimental results have confirmed that a two-level system interacting with a single mode can exhibit energy splitting renormalization of up to 15\%, with a time-dependent profile consistent with the predictions of the effective Hamiltonian obtained from the coherent part of the system's TCL master equation~ \cite{Colla2025exp}. Furthermore, the effective Hamiltonian has been proposed as a key operator in the generalization of the laws of quantum thermodynamics to the strong-coupling regime~\cite{Colla2022, Picatoste2024, Colla2024roles}.

In summary, we present a recursive perturbative expansion for the TCL generator,
formulated in terms of left- and right-acting operators, and use this approach to obtain a systematic
expression for the generator in a generalized Lindblad form.
Our formulation does not impose constraints on the system-bath structure beyond an initially uncorrelated state, making it broadly applicable to a wide range of non-Markovian quantum systems.

The remainder of the paper is structured as follows: Sec.~\ref{sec:expansion} details the recursive perturbative expansion of the TCL master equation in terms of left-right operators. Sec.~\ref{sec:structure} introduces the canonical decomposition of the generator into its coherent and dissipative components, with a focus on traceless Lindblad operators and the role of minimal dissipation. Finally, we apply this method in Sec.~\ref{sec:Ks-expansion} to present an explicit expansion up to fourth order.

\section{Recursive perturbation expansion of the generator}\label{sec:expansion}
\subsection{Assumptions}
We consider a general bipartite quantum system, where one part of the bipartition corresponds to the system of interest \( S \) and the other to its environment \( E \). To describe the effective dynamics of \( S \), we trace out the degrees of freedom of \( E \), obtaining the reduced evolution of the system. 

We assume that the total Hamiltonian governing the evolution of the combined system is given by
\begin{align}
H_t = H_{S,t} + H_{E,t} + \lambda A^{\mathcal{S}}_t \otimes B^{\mathcal{S}}_t \;,
\end{align}
where \( H_{S,t} \) and \( H_{E,t} \) describe the intrinsic dynamics of the system and the environment, respectively. The interaction between them is characterized by the coupling parameter \( \lambda \), with \( A^{S}_t \) and \( B^{S}_t \) being time-dependent operators acting on the Hilbert spaces of the system and the environment, respectively.

We assume the interaction term to be in a tensor product form and that \( A^\dagger = A \), \( B^\dag = B \) for simplicity and clarity. 
However, the calculations can be easily extended to a more general interaction of the form \( \lambda \sum_i A_{i,t} \otimes B_{i,t} \). The subscript \( t \) indicates the explicit time dependence of all terms in \( H_t \), which allows us to describe bipartite systems that are not isolated but still evolve unitarily. The superscript \( \mathcal{S} \) indicates that the operator is written in the Schrödinger picture.

The total initial state is assumed to be uncorrelated, i.e. ${\rho}_{SE}(0)=\rho_S\otimes \rho_E$. 
If initial correlations are present, one may define the correlation matrix
$\chi := {\rho}_{SE}(0)-\Tr_{S}\{{\rho}_{SE}(0)\}\otimes \Tr_{E}\{{\rho}_{SE}(0)\}$ and treat \(\chi\) as a known contribution to the initial conditions for the irrelevant degrees of freedom. 
This approach yields an affine, yet linearizable, dynamical map for the reduced system~\cite{Stelmachovic2001,Colla2022corr}.

We move into the interaction picture by evolving the bipartite system under the free local evolution given by $H_S$ and $H_E$. We denote the interaction picture Hamiltonian of system and environment by $\tilde{H}_t= \lambda A_t\otimes B_t$ and the interaction picture density matrix by $\tilde\rho_{SE}(t)$, so that the total evolution is given by $\dot{\tilde{\rho}}_{SE}(t) = -i [\tilde{H}_t, \tilde{\rho}_{SE}(t)]$. Under these assumptions, the total unitary evolution of the bipartite system can be described by the following expression
\begin{align}
\mathcal{U}_t[{\rho}_{SE}(0)]= \mathcal{T}\left( e^{-i \int_0^t d\tau(\tilde{H}_{\tau}^L - \tilde{H}_{\tau}^R )}\right)\rho_S\otimes \rho_E \; ,
\end{align}
where $\mathcal{T}$ represents time-ordering and the superscripts $L$ and $R$ denote left-acting and right-acting (super)operators respectively, namely:
\begin{align}
X^L[\rho] &= X\rho, \;\hspace{1cm}  
X^R[\rho] = \rho X \; ,
\end{align}
where both $X$ and $\rho$ are operators.

\subsection{Expansion of the reduced dynamical map}

The dynamical map describing the evolution of the reduced system is obtained by tracing out the environmental degrees of freedom: 
\begin{align} 
    \Phi_t[\rho_S] = \Tr_E{\mathcal{U}_{t}[\rho_S\otimes \rho_E]}. 
\end{align}
To expand this map in powers of the coupling strength $\lambda$, we express the unitary evolution operator as a Taylor series, yielding 
\begin{align} 
    \Phi_{t} = 1 + \sum_{n=1}^{\infty} (-i \lambda )^n \mu_{n}, 
\end{align} 
where each term $\mu_{n}$ is time-dependent (we suppress the explicit subscript $t$ for clarity) and given by 
\begin{align}
    \mu_n [\rho_S] = \frac{1}{n!} \Tr_E \left\{ \mathcal{T}\left(\int_0^t d\tau \left(\tilde{H}_{\tau}^L - \tilde{H}_{\tau}^R \right) \right)^n \rho_S\otimes \rho_E\right\} \;.
\end{align}    
We can now make the dependency on the system and bath operators $A$ and $B$ explicit. We make use of the binomial theorem, and recognize that left- and right- acting operators can be separately time ordered, as they commute at all times, to obtain
%\begin{widetext}
\begin{align}\label{eq:moment}
\mu_n [\rho_S]=&  \sum_{k=0}^n \frac{(-1)^{n-k}}{(n-k)!k!}  \Tr_E \left\{ \mathcal{T}\left(\int_0^t d\tau A_{\tau}^L B_{\tau}^L\right)^{k}\right. \nonumber\\ &\hspace{2cm}\left.\mathcal{T}\left( \int_0^t d\tau A_{\tau}^R B_{\tau}^R \right)^{(n-k)}\hspace{-0.5cm} \rho_S\otimes \rho_E\right\} \; .
\end{align}
%\end{widetext} 
The idea is now to separate the action of $A$ and $B$ on $\rho_S\otimes\rho_E$ in order to easily perform the partial trace with respect to the environment. We however have to keep track of the ordered time-dependencies. We thus define for left- and right-acting superoperators the notation
\begin{align}
{A^{L/R}}(\boldsymbol{t}_1^k)[\,\cdot\,] :=& A^{L/R}_{t_1}\circ{A^{L/R}}_{t_2}\circ...\circ{A^{L/R}}_{t_k}[\,\cdot\,] \\
{A^{L/R}}^\dag(\boldsymbol{t}_1^k)[\,\cdot\,] :=& A^{L/R}_{t_k}\circ{A^{L/R}}_{t_{k-1}}\circ...\circ{A^{L/R}}_{t_1}[\,\cdot\,]
\end{align}
and analogously for products of operators
\begin{align}
{A}(\boldsymbol{t}_1^k)&:= A_{t_1}\cdot{A}_{t_2}\cdot...\cdot{A}_{t_k} \\
{A}^\dag(\boldsymbol{t}_1^k)&:= A_{t_k}\cdot{A}_{t_{k-1}}\cdot...\cdot{A}_{t_1} \; .
\end{align}
With this notation we can rewrite each contribution $\mu_n$, eq.~\eqref{eq:moment}, as a sum of terms 
\begin{align}
\mu_n &=\sum_{k=0}^{n}\mu_{n}^{k}
\end{align}
where each contribution is given by
\begin{align}
\mu_{n}^{k}= (-1)^{n-k} \int_0^t d \boldsymbol{\tau}_{1}^{k} d \boldsymbol{s}_{1}^{n-k}  & A^L(\boldsymbol{\tau}_1^k)  A^R(\boldsymbol{s}_{1}^{n-k})\nonumber \times \\ & \times D(\boldsymbol{\tau}_{1}^{k},\boldsymbol{s}_{1}^{n-k}) \; , \label{eq:mu_kn}
\end{align}
which contains a product of $k$ left-acting operator times $n-k$ right-acting operators.
For the sake of readability, we have introduced a few more notation conventions. The first is the integral notation, which is made to mean the following:
\begin{align}\label{eq:time_integrals}
\int_0^t d \boldsymbol{\tau}_{1}^{k} d \boldsymbol{s}_{1}^{n-k} \; (...)\; = &\int_0^t d \tau_{1} \int_0^t d\tau_2 \; ... \int_0^t d\tau_{k} \times \\ &\times \int_0^t d s_{1} \int_0^t d s_2 \;... \int_0^t ds_{n-k} \; (...)\; . \nonumber
\end{align}
Then, we have defined the coefficients $D$, which include the environmental correlation functions (at various orders) as well as the time-ordering. Explicitly, they read
\begin{align}\label{eq:D1}
D(\boldsymbol{\tau}_{1}^{k},\boldsymbol{s}_{1}^{n-k})=& \Tr_E \left\{B^{\mathrm{R}}(\boldsymbol{s}_{1}^{n-k})\circ B^{\mathrm{L}}(\boldsymbol{\tau}_{1}^{k})[\rho_{E}] \right\}\theta_{\boldsymbol{\tau}_{1}^{k}}\theta_{\boldsymbol{s}_{1}^{n-k}} \;,
\end{align}
where the $\theta_{\boldsymbol{\tau}_{1}^{n}}$-functions is defined as
\begin{align}
\theta_{\boldsymbol{\tau}_{1}^{n}} = 
\begin{cases}
1, & \text{if } \tau_1 > \tau_2 > \cdots > \tau_n \\
0, & \text{otherwise}
\end{cases}
\end{align}
so that
\begin{align}
    \int_{0}^{t} d\boldsymbol{\tau}_{1}^{n}\,\theta_{\boldsymbol{\tau}_{1}^{n}} 
= \int_{0}^{t} d\tau_1 \int_{0}^{\tau_1} d\tau_2 \cdots \int_{0}^{\tau_{n-1}} d\tau_n.
\end{align}
Furthermore, notice that the function $D$ satisfies 
\begin{align}\label{eq:D_D}
D^*(\boldsymbol{\tau}_{1}^{k},\boldsymbol{s}_{1}^{n-k})= D(\boldsymbol{s}_{1}^{n-k},\boldsymbol{\tau}_{1}^{k})\; .\end{align}

\subsection{Expansion of the time-local generator}

The exact time-local generator for the reduced system is given by composing the derivative of the dynamical map with its inverse \cite{Hall2014,Breuer2012}, i.e. $\Lt_t = \dot\Phi_t\circ \Phi_t^{-1}$. 
If $\norm{\Phi_{t}-\text{id}} < 1$~\footnote{here $\norm{\cdot}$ must be intended as the operator norm}, then $\Lt_t$ can be then written as a combination of the terms $\mu_n^k$ and their time derivatives. Those can in turn be rearranged, in order to collect different terms according to the exponent of the coupling coefficient $\lambda$. This procedure was already performed in Ref.~\cite{Gasbarri2018} for an analogous expansion. Following the steps within, one obtains the expansion of the generator in terms of $\lambda$:
\begin{align}\label{eq:lser}
    \Lt_t = \sum_{n=0}^{\infty}\lambda^n \Lt_n \; ,
\end{align}
where again we have dropped the subscript $t$ from the $n$-th order term of the generator, which is however still time-dependent and is given by
\begin{align}\label{eq:gen-exp-lambda}
\mathcal{L}_{n} &= (-i)^n \sum_{q=0}^{n-1} (-1)^q \hspace{-2em} \sum_{(m_0+...+m_q=n)} \sum_{k_{0}=0}^{m_{0}}\sum_{k_{q}=0}^{m_{q}}\dot{\mu}_{m_0}^{k_{0}} \mu_{m_1}^{k_{1}}... \mu_{m_q}^{k_{q}}\nonumber\\
&= (-i)^n\sum_{k=0}^{n} \sum_{q=0}^{n-1} (-1)^q \hspace{-2em} \sum_{(m_0+...+m_q=n)}\sum_{(k_{0}+\dots + k_{q} = k)} \hspace{-1.5em} \dot{\mu}_{m_0}^{k_{0}} \mu_{m_1}^{k_{1}}... \mu_{m_q}^{k_{q}}\nonumber\\
&= \sum_{k=0}^{n} \mathcal{L}_{n}^{k} \; .
\end{align}
In the above, $\dot{\mu}_{m}^{k}$ represents the first time derivative of ${\mu}_{m}^{k}$, which can be expressed as
\begin{align}\label{eq:mu_kn_dot}
\dot{\mu}_n^{k} &=  \int_0^t d \boldsymbol{\tau}_{1}^{k} d \boldsymbol{s}_{1}^{n-k}  A^L(\boldsymbol{\tau}_{1}^k) A^R(\boldsymbol{s}_{1}^{n-k}){\dot{D}(\boldsymbol{\tau}_{1}^{k},\boldsymbol{s}_{1}^{n-k})} \; ,
\end{align}
with the time derivative of the bath correlation functions given by
\begin{align}
\dot{D}(\boldsymbol{\tau}_{1}^{k},\boldsymbol{s}_{1}^{n-k}):=D(\boldsymbol{\tau}_{1}^{k},\boldsymbol{s}_{1}^{n-k})
 \big( \delta_{\tau_1,t}+\delta_{s_{1},t} \big) \; ,
\end{align}
since $\tau_1$ and $s_{1}$ are the only largest times in each term due to time-ordering. Notice that in the above we have assumed the following convention for each $\delta$-function: $\int_0^t d\tau \delta_{\tau,t} f(\tau) = f(t)$ for any function $f$.

In the last line of eq.~\eqref{eq:gen-exp-lambda}, each term $\mathcal{L}_{n}^{k}$ describes the $n$-th element of the series that contains a string of $k$ left-acting operators. Its explicit form can be obtained by replacing each $\mu_{m_{i}}^{k_{i}}$ with their explicit formula and relabelling in increasing order the time integrals. Collecting all coefficients together, one then obtains the expression for each term $\mathcal{L}_{n}^{k}$ in terms of left and right-acting operators, namely
\begin{align}
\mathcal{L}_{n}^{k} =  i^n(-)^{k}\int_0^t d \boldsymbol{\tau}_{1}^{k} d \boldsymbol{s}_{1}^{n-k} \mathcal{D}(\boldsymbol{\tau}_{1}^{k},\boldsymbol{s}_{1}^{n-k}) A^{L}(\boldsymbol{\tau}_1^k) A^{R}(\boldsymbol{s}_{1}^{n-k}) \;. \label{eq:Lkn}
\end{align}
In the above, we have defined the environmental generalized cumulants, which read:
\begin{align}\label{eq:Dcal}
  &\mathcal{D}(\boldsymbol{\tau}_{1}^{k},\boldsymbol{s}_{1}^{n-k})\nonumber\\
  &\hspace{0.1cm}=\sum_{q=1}^{n}(-)^{q+1}\hspace{-0.9cm}\sum_{\substack{ 0 \le k_{1}\le ...\le k_{q}= k \\
     0\le m_{1}\le ...\le m_{q}= n-k}}\hspace{-1cm}
    \dot{D}(\boldsymbol{\tau}_{1}^{k_{1}},\boldsymbol{s}_{1}^{m_{1}})\prod_{j=1}^{q-1} D(\boldsymbol{\tau}_{1+k_{j}}^{k_{j+1}},\boldsymbol{s}_{1+m_{j}}^{m_{j+1}}) \; .
\end{align}

Equation~\eqref{eq:Dcal} is general for all terms and quite cumbersome. However, several terms $D$ appearing in it will often not be present in practice, as we have assumed the following rule of notation:
\begin{equation}  D(\tau_{a_1}^{b_1},s_{a_2}^{b_2}) = \begin{cases} D(\tau_{a_1}^{b_1},s_{a_2}^{b_2}) \quad &\text{for } b_1\geq a_1 \land b_2\geq a_2 \\
D(\tau_{a_1}^{b_1}) &\text{for } b_1\geq a_1 \land b_2< a_2 \\
D(s_{a_2}^{b_2}) &\text{for } b_1< a_1 \land b_2\geq a_2 \\
0 &\text{for } b_1 < a_1 \land b_2 < a_2
\end{cases} \; 
\end{equation}
which can also be summarized as the following:
\begin{align}\label{eq:Dcal-rule}
D(\tau_{a_1}^{b_1},s_{a_2}^{b_2}) =& D(\tau_{a_1}^{b_1},s_{a_2}^{b_2}) \theta_{a_1,b_1}\theta_{a_2,b_2} \nonumber \\ &+ D(\tau_{a_1}^{b_1}) \theta_{a_1,b_1}(1-\theta_{a_2,b_2}) \nonumber \\ & + D(s_{a_2}^{b_2}) (1-\theta_{a_1,b_1})\theta_{a_2,b_2} \; ,
\end{align}
where $\theta_{a,b}$ is a discrete theta function (with convention $\theta_{a,b}= 1$ for $a\le b$).

Keep in mind that the times in the vectors $\boldsymbol{\tau}_{1}^{k}$ and $\boldsymbol{s}_{1}^{n-k}$ appearing in eq.~\eqref{eq:Dcal} are not actually ordered in a particular direction, as the $\theta$ functions appearing in $ \mathcal{D}(\boldsymbol{\tau}_{1}^{k},\boldsymbol{s}_{1}^{n-k})$ are responsible for the time ordering of all operators and functions in the integrals. We have named them generalized cumulants as they share formal similarities to the ordered cumulants of van Kampen \cite{VanKampen1974a,VanKampen1974b}.

Each term in equation \eqref{eq:Dcal} can be obtained recursively if all the terms of lower $n$ are known. We can see this by rearranging eq.~\eqref{eq:Dcal} in order to highlight lower order contributions (see Appendix~\ref{sec:appA}). This gives
\begin{align}\label{eq:Dcal-rec}
  \mathcal{D}(\boldsymbol{\tau}_{1}^{k},\boldsymbol{s}_{1}^{n-k}) =&
  \dot{D}(\boldsymbol{\tau}_{1}^{k},\boldsymbol{s}_{1}^{n-k}) \nonumber \\ &-\sum_{l=0}^{k}\sum_{r=0}^{n-k}   \mathcal{D}(\boldsymbol{\tau}_{1}^{l},\boldsymbol{s}_{1}^{r}){D}(\boldsymbol{\tau}_{l+1}^{k},\boldsymbol{s}_{r+1}^{n-k}) \; ,
\end{align}
where, analogously as before, some terms may drop out due to the rule of notation \eqref{eq:Dcal-rule}.

As we will see later in Sec.~\ref{sec:Ks-expansion}, the recursive expression for the cumulants is particularly helpful to compute contributions at higher orders, especially whenever the system is such that some lower order cumulants simplify or vanish entirely.
Combining eqs.~\eqref{eq:gen-exp-lambda}, \eqref{eq:Lkn} and \eqref{eq:Dcal-rec}, we find that $n$-th order contribution to the generator of the dynamics acts as the following on any operator:
\begin{align}
\Lt_n [X] =  (i)^n \sum_{k=0}^n(-)^{k}\int_0^t & d \boldsymbol{\tau}_{1}^{k} d \boldsymbol{s}_{1}^{n-k} \mathcal{D}(\boldsymbol{\tau}_{1}^{k},\boldsymbol{s}_{1}^{n-k}) \times \nonumber \\ & \times A(\boldsymbol{\tau}_1^k) X A^\dag(\boldsymbol{s}_{1}^{n-k}) \; . \label{eq:Ln}
\end{align}
The term described by this expression is analogous to eq.~(16) in~\cite{Gasbarri2018}, but here expressed in terms of left- and right-acting operators. 
This representation, unlike the expression found in~\cite{Gasbarri2018}, allows the series to be rewritten in a canonical form, which enables the distinction between the coherent and incoherent contributions to the dynamics, as we will discuss in the next section.

\section{Canonical representation of the Time-Local Generator}\label{sec:structure}

The time-local generator governing the reduced dynamics of an open quantum system can always be expressed in the canonical form:
\begin{align}
\mathcal{L} [X] =& -i [K, X]  \nonumber \\ \label{eq:gen-lindblad-str}
&+ \sum_{ij} \gamma_{ij} \left( L_{i} X L_{j}^{\dagger} -\frac{1}{2} \left\{ L_{j}^{\dagger} L_{i}, X \right\} \right) \;.
\end{align}
Here, all contributions are explicitly time-dependent, and the coefficient matrix $\gamma_{ij}$ is Hermitian. The structure of this generator ensures that the dynamics preserve Hermiticity and trace but do not necessarily guarantee complete positivity~\cite{Hall2014, Breuer2012}. If $\gamma_{ij}$ remains positive semi-definite at all times, then the evolution is CP-divisible, corresponding to a Markovian process. However, when some eigenvalues of $\gamma_{ij}$ become negative, the dynamics exhibit non-CP-divisibility, a possible signature of non-Markovianity~\cite{Rivas_2014,Breuer2016C}.

The coherent part of the generator, determined by the effective Hamiltonian $K$, governs unitary-like evolution. Importantly, $K$ differs from the bare system Hamiltonian $H_S$ due to renormalization effects induced by system-environment interactions. The dissipative part describes irreversible evolution of the system state and is characterized by the Lindblad operators $L_i$ and the time-dependent rates $\gamma_{ij}$. 

The possibility to recast the perturbation series described by~\eqref{eq:Ln} into the general form of eq.~\eqref{eq:gen-lindblad-str} facilitates a physical interpretation of its individual components, enabling a systematic distinction between coherent and dissipative contributions. However, this decomposition is inherently non-unique, introducing an additional layer of complexity.

\subsection{Ambiguity of the effective Hamiltonian and the principle of minimal dissipation}

 It is well known that in a standard Lindblad master equations, the Hamiltonian term can include not only the Lamb shift but also additional contributions arising from invariance transformations~\cite{Breuer2007}.

For a generalized time-local generator in eq.~\eqref{eq:gen-lindblad-str}, the following transformations preserve the overall structure but modify the explicit representation of the coherent and dissipative dynamics:
\begin{align}\label{eq:inv1}
L_i &\longrightarrow L_i - \alpha_i \mathbb{1} \; , \\
K &\longrightarrow K + \frac{1}{2i} \sum_{ij} \gamma_{ij} \left( \alpha_i L_j^\dag -\alpha_j^* L_i \right) \; . \label{eq:inv2}
\end{align}
These transformations show that the effective Hamiltonian $K$ can acquire additional terms that depend explicitly on the Lindblad operators $L_{i}$ and dissipation rates $\gamma_{ij}(t)$.

Moreover, any purely Hamiltonian contribution to $K$ (involving only system degrees of freedom) can always be rewritten in a dissipator-like form~\cite{Chruscinski2022, Sorce2022}, further emphasizing the inherent ambiguity in defining $K$.
 While the generator $\mathcal{L}$ uniquely determines the system's dynamics, the specific division between coherent and dissipative contributions remains representation-dependent.

Given this ambiguity, an additional constraint is needed to define a physically meaningful decomposition of the generator in terms a coherent and dissipative component. One approach is the principle of minimal dissipation, which minimizes the dissipative contribution of $\mathcal{L}$ using a norm that averages over random input and output states~\cite{Sorce2022,Colla2022}.
 It has been shown that this criterion is equivalent to writing the dissipator using traceless Lindblad operators, thereby recovering the standard Lamb-shifted Hamiltonian in the CP semigroup case. By enforcing this condition, one obtains a well-defined representation of the generator.

In the next section, we reformulate the microscopic generator derived in Sec.~\ref{sec:expansion} into a generalized Lindblad form satisfying the minimal dissipation principle. 

\subsection{Perturbative structure of the generator}
The generator defined via the perturbative expansion in eq.~\eqref{eq:lser}, eq.\eqref{eq:Ln} and eq.~\eqref{eq:Dcal-rec} exhibits the following general structure  
\begin{align}\label{eq:gen-LR}
\mathcal{L}[X]= \sum_{ij} \omega_{ij} V_i^{L} (V_j^{\dagger}) ^{R}[X]= \sum_{ij} \omega_{ij} V_i X V_j^{\dagger} \;,
\end{align}
where $\omega_{ij}^{*} =\omega_{ji}$ (see eq.~\eqref{eq:D_D}) and where the set of operators $\{V_i\}$ may be overcomplete or linearly dependent.
Without loss of generality, we label $V_{0}= \mathbb{1}$.

We recall that the TCL generator satisfies the conditions of Hermiticity preservation and trace annihilation, i.e. 
\begin{align}\label{eq:LR-Herm-pres}
    &\sum_{ij} \omega^*_{ij} (V_j)^{L} (V_i^{\dag}) ^{R} = \sum_{ij} \omega_{ij} V_i^{L} (V_j^{\dag} )^{R}, \\
    &\sum_{ij} \omega_{ij} V_j^{\dag} V_i =0 \;. \label{eq:trace-annihilation}
\end{align}
From trace annihilation, it is not difficult to obtain 
\begin{align}\label{eq:scatt}
  \sum_{i,j \neq 0} \omega_{ij}V_{j}^{\dag}V_{i} = -\sum_{i\neq 0}(\omega_{i0} V_{i} +\omega_{0i}V_{i}^{\dag}) \; .
\end{align}
One may notice the resemblance of this equation to the unitarity condition commonly used in scattering theory to derive the optical theorem~\cite{reed1979iii, taylor2012scattering}. Similarly, in collisional dynamics, such conditions are employed to derive Lindblad-like master equations when interactions are described via a scattering operator~\cite{diosi1995quantum,hornberger2003collisional,hornberger2008monitoring}.
Moreover, using the identity 
\begin{align}
\omega_{i0} V_i X +\omega_{0i} X V^\dag_i &=  \frac{1}{2} \{\omega_{i0} V_i  +\omega_{0i}  V^\dag_i, X\} \nonumber \\ 
&+ \frac{1}{2} [ \omega_{i0} V_i - \omega_{0i} V^\dag_i,X] \; ,
\end{align}
along with eq.~\eqref{eq:scatt}, one can rewrite the generator in a generalized Lindblad form:
\begin{align}\label{eq:GGorini}
\mathcal{L}[X]=& -i \sum_{i}\left[\frac{1}{2i} (\omega_{0i} V^\dag_i -  \omega_{i0} V_i ),X\right] \nonumber \\ 
&+\sum_{ij\neq 0} \omega_{ij}\left( V_i X V_j ^{\dag} -\frac{1}{2}  \{ V_j^{\dagger}V_i,X\} \right). 
\end{align}
This result generalizes the work of Gorini et al.~\cite{Gorini1976a} by extending it to non-Markovian dynamics and allowing for arbitrary sets of operators $V_{i}$.
However, in the general case, the Hamiltonian-like contribution is guaranteed to be Hermitian only if the condition $\omega^{*}_{0i} =\omega_{i0}$ is satisfied. 
This condition ensures that the effective Hamiltonian remains a valid observable and prevents the introduction of spurious non-Hermitian terms. Notably, this symmetry property holds naturally within the perturbative expansion considered in this work (see eq.~\eqref{eq:D_D}).
Equation~\eqref{eq:GGorini} allows to rewrite the $n$-th term of the series described by eq.~\eqref{eq:Ln}  as 
  \begin{widetext}
\begin{align}
 \mathcal{L}_{n}[X] &= -i\int_0^t d \boldsymbol{s}_{1}^{n} \left[\text{Im}\left\{ (i)^n\mathcal{D}(\boldsymbol{s}_{1}^{n}) {A}(\boldsymbol{s}_{1}^{n})^{\dagger}\right\},X\right] \nonumber\\
  &\hspace{1cm}+\sum_{k=0}^{n}(-)^k i^n \int_0^t d \boldsymbol{\tau}_{1}^{k} d\boldsymbol{s}_{1}^{n-k} \mathcal{D}(\boldsymbol{\tau}_{1}^{k},\boldsymbol{s}_{1}^{n-k}) \left({A}(\boldsymbol{\tau}_{1}^{k}) X {A}(\boldsymbol{s}_{1}^{n-k})^{\dagger}-\frac{1}{2}\left\{ {A}(\boldsymbol{s}_{1}^{n-k})^{\dagger}{A}(\boldsymbol{\tau}_{1}^{k}),X\right\} \right). \label{eq:gen-lindblad0}
  \end{align}
  \end{widetext}

However, it is clear that the choice of the operator set influences the form of the Hamiltonian contribution.
Inspired by scattering theory, where the transition operator is taken to be traceless, or in a different manner by minimal dissipation principle (see discussion above), we impose the further constraint of traceless Lindblad operators.
This is achieved by applying the transformation
\begin{align}
V_{i} \to \overline{V}_{i}\equiv V_{i}-\braket{V_{i}}_{1/d}
\end{align}    
where $\braket{V_{i}}_{1/d}$ referes to the average over the maximally mixed state $\mathbb{1}/d $. This transformation, together with the invariance transformations in eqs.~(\ref{eq:inv1}, \ref{eq:inv2}), leads to the following structure for the generator:
\begin{align}
    \mathcal{L}[X]=& -i [K_S,X] \nonumber \\ &+\sum_{ij}\omega_{ij}\left( \overline{V}_i X \overline{V}_j ^{\dag} - \frac{1}{2}\{ \overline{V}_j^{\dagger}\overline{V}_i,X\} \right).
    \end{align}
  where the effective Hamiltonian is given by 
  \begin{align}
  K_{S} = \frac{1}{2i} \sum_{ij} \omega_{ij} \left( \braket{V_{i}}_{1/d}V^\dag_{j}- \braket{V^\dag_{j}}_{1/d}V_{i}\right)\; .
  \end{align}
  Due to eq.~\eqref{eq:LR-Herm-pres}, the effective Hamiltonian is always Hermitian, and can be equivalently rewritten as
  \begin{align}
  K_{S} = \sum_{ij} \text{Im}\left\{ \omega_{ij} \braket{V_{i}}_{1/d}V^\dag_{j} \right\}\; .
  \end{align}
  This formulation leads to a well-defined expression for the effective Hamiltonian, even if $\omega_{0i}^{*}\neq \omega_{i0}$, providing a systematic and straightforward method for deriving a Lindblad-like generator that preserves both Hermiticity and trace annihilation.

Using this strategy, we can express each order $(\mathcal{L}_{n})$ of the pertubative series for the TCL generator $\mathcal{L}_{t} = \sum_{n=1}^{\infty}\lambda^{n}\mathcal{L}_{n}$ in a generalized Lindblad form with traceless Lindblad operators, i.e.
  \begin{widetext}
\begin{align}
 \mathcal{L}_{n}[X] &= \sum_{k=0}^{n}(-)^k\int_0^t d \boldsymbol{\tau}_{1}^{k} d\boldsymbol{s}_{1}^{n-k} \Bigg\{-i\left[\text{Im}\left\{ (i)^n\mathcal{D}(\boldsymbol{\tau}_{1}^{k},\boldsymbol{s}_{1}^{n-k})\langle A(\boldsymbol{\tau}_{1}^{k})\rangle_{1/d} {A}(\boldsymbol{s}_{1}^{n-k})^{\dagger}\right\},X\right] \nonumber\\
  &\hspace{4cm}+i^n\mathcal{D}(\boldsymbol{\tau}_{1}^{k},\boldsymbol{s}_{1}^{n-k}) \left(\overline{A}(\boldsymbol{\tau}_{1}^{k}) X \overline{A}(\boldsymbol{s}_{1}^{n-k})^{\dagger}-\frac{1}{2}\left\{ \overline{A}(\boldsymbol{s}_{1}^{n-k})^{\dagger}\overline{A}(\boldsymbol{\tau}_{1}^{k}),X\right\} \right)\Bigg\}, \label{eq:gen-lindblad}
  \end{align}
  \end{widetext}
where $\bar{A}$ is the null trace operator defined as $\bar{A} = A-\braket{A}_{1/d}\mathbb{1}$.
With this formulation in place, we now turn our attention to the Hamiltonian part of the generator, examining its structure and perturbative expansion.

\subsection{Perturbative expansion of the effective Hamiltonian}\label{sec:Ks}
The structure of the effective system Hamiltonian naturally emerges from eq.~\eqref{eq:gen-lindblad} as a perturbative series in powers of the coupling strength $\lambda$, i.e.
\begin{align}
    K(t) = \sum_{n=0}^{\infty} \lambda^n K_n \; 
\end{align}
with
\begin{align}\label{eq:Kn}
     K_n
    = \frac{-(+i)^n}{2i}&\sum_{k=0}^n (-)^{k} \int_0^t d \boldsymbol{\tau}_{1}^{k} d \boldsymbol{s}_{1}^{n-k}\times  \\ \nonumber & \times \left[ \mathcal{D}(\boldsymbol{\tau}_{1}^{k},\boldsymbol{s}_{1}^{n-k})\Xcal{\mathbb{A}}{1}{k}{1}{n-k} - (-)^{n} \text{h.c.}\right] \; ,
\end{align}   where, for later convenience, we have defined the operator $\Xcal{\mathbb{A}}{1}{k}{1}{n-k} := \langle A(\boldsymbol{s}_{1}^{n-k})^{\dagger}\rangle_{1/d}  A(\boldsymbol{\tau}_{1}^{k})$.

This equation provides a systematic method for computing each perturbative term of the emergent Hamiltonian directly from the underlying system-environment microscopic model. The recursive nature of eq.~\eqref{eq:Dcal-rec}, as previously mentioned, often simplifies calculations, particularly when the environmental generalized cumulants exhibit special properties, such as vanishing contributions.

On the other hand, the explicit structure of eq.~\eqref{eq:Kn} highlights the operatorial structure of $K$, revealing its dependence on the system-side interaction operator. This structure simplifies significantly when the dynamics lead to pure dephasing (namely when $[H_S,A]=0$), as $A(\boldsymbol{\tau}_{1}^{k})= A^k$. Moreover, for finite-dimensional systems (particularly at lower dimensions) one can exploit algebraic properties of the operators to infer possible features of each contribution (see, for example, the discussion on spin systems in~\cite{Colla2025K}).

To gain further insight into the structure of the effective Hamiltonian, we define the following operator:
\begin{align}\label{eq:Knk}
K_n^k = (-)^{k} \int_0^t d \boldsymbol{\tau}_{1}^{k} d \boldsymbol{s}_{1}^{n-k} \mathcal{D}&(\boldsymbol{\tau}_{1}^{k},\boldsymbol{s}_{1}^{n-k})\Xcal{\mathbb{A}}{1}{k}{1}{n-k} \; ,
\end{align}
This expression represents the partial contribution to $K$ at order $n$, specifically arising from the product of $k$ interaction-picture operators $A$. The full $n$-th order contribution to $K$ can then be written as
\begin{align}\label{eq:Kn-short}
    K_n = \frac{-(+i)^n}{2i}\sum_{k=0}^n\left[ K_n^k - (-)^{n} \text{h.c.}\right] \; ,
\end{align}
This expression reveals that the contributions to $K$ take different forms depending on whether $n$ is even or odd, specifically:
\begin{align}\label{eq:Kn-short-even}
    K_{2m} &= (-1)^{m+1}\sum_{k=0}^{2m}\text{Im}\left(K_{2m}^k\right) \; , \\
    \label{eq:Kn-short-odd}
    K_{2m+1} &= (-1)^{m+1}\sum_{k=0}^{2m+1}\text{Re}\left(K_{2m+1}^k\right) \; .
\end{align}
Thus, for even-order terms only the imaginary part of $K_n^k$ contributes to the Hamiltonian, while for odd-order terms, only the real part does. This distinction plays an important role in understanding how different orders of the perturbative expansion affect the system’s effective dynamics. For further details on the topic and a discussion on the role of fluctuations and dissipation on the Hamiltonian term see~\cite{Colla2025K}.

In the next section, we provide explicit calculations of each term up to third order, as well as the fourth-order contribution in cases where the lower-order terms vanish. 

\section{Explicit expansion up to fourth order}\label{sec:Ks-expansion}
In this section, we exploit our method -- namely, eqs.~\eqref{eq:Dcal} or \eqref{eq:Dcal-rec} (recommended) for the coefficients -- to calculate the explicit form of the generalized cumulants. We then apply these results to eq.~\eqref{eq:Kn}  to obtain an explicit expression for the effective Hamiltonian up to low orders of perturbation. The dissipative part of the dynamics is trivially obtained once the form of the cumulants is given (see eq.~\eqref{eq:gen-lindblad}).
We perform the procedure for the most general case until third order, while we exploit in particular eq.~\eqref{eq:Dcal-rec} to calculate the fourth order contribution under the assumption of vanishing first order.
Notice that the results of this section can be readily used to obtain the perturbative expansion of the Hamiltonian term in eq.~\eqref{eq:gen-lindblad0} by replacing $\langle A(\boldsymbol{\tau}_{1}^{n-k})\rangle_{1/d}  \to 0$.

\subsection{First order}
At first order there are only two possible coefficients which are simply
\begin{align}
\Dcals{1}{} &= \Dfds{1}{} \; , \\
\Dcalt{1}{} &= \Dfdt{1}{}\; .
\end{align}
Inserted into \eqref{eq:Kn}, they give the following contribution to the effective Hamiltonian:
\begin{align}
    K_1 = \braket{B_t}\left( A_t - \braket{A_t}_{1/d}\right) \; .
\end{align}
Most often, the interaction operator $A$ is taken to be traceless -- if not, the contribution responsible for it can be absorbed into the environment's bare Hamiltonian. In this case, the Schr\"{o}dinger picture contribution to $K$ is easily found to be $K_1^{\mathcal{S}}(t) = \braket{B_t}A$.
It represents driving on the system degrees of freedom due to a non-vanishing $\braket{B_t}$. Furthermore, a commutator with this Hamiltonian is the only first-order contribution in the expansion of the generator, while there are no dissipative contributions:
\begin{align}
    \Lt_1 = -i [K_1, \cdot] \; .
\end{align}

This can be interpreted as the emergence of effective driving on the system, which is often due to coherences in the initial state of the environment. The same expression for this contribution can be obtained using known techniques in quantum optics, for example~\cite{Cohen1998}. Moreover, in certain models this contribution is the only emergent driving term, with negligible dissipation, that survives under a suitable semiclassical limit (weak coupling $\lambda$ and large $\braket{B_t}$)\cite{Colla2024roles}.

\subsection{Second order}
At second order, the coefficients $\mathcal{D}$ are given by two-point correlation functions of the bath. From the different time-dependencies, there are three different coefficients which read
\begin{align}
&\Dcals{1}{2} = \Dfds{1}{2} -\Dfds{1}{}\Dfs{2}{} \; ,\\
&\Dcal{1}{}{1}{} = \Dfd{1}{}{1}{} - \Dfdt{1}{}\Dfs{1}{} \nonumber\\
&\quad\quad \quad\quad \quad\quad - \Dfds{1}{}\Dft{1}{} \; ,\\
&\Dcalt{1}{2} = \Dfdt{1}{2} -\Dfdt{1}{}\Dft{2}{} \; .
\end{align}

Inserted into the expression for $K$, they give the following second order contribution to the Hamiltonian:
\begin{align}
    K_2 = \frac{1}{2i} \int_0^t dt_1 &\left( \braket{B_t B_{t_1}} - \braket{B_t}\braket{B_{t_1}} \right) \times \nonumber \\ &\times \big[ A_t A_{t_1} - \braket{A_t A_{t_1}}_{1/d} \\ &\quad+ A_t \braket{A_{t_1}}_{1/d} - A_{t_1} \braket{A_t}_{1/d} \big] - \text{h.c.} \; .
\end{align}
In the common case where the system operator $A$ in the interaction Hamiltonian is traceless, and the bath operator $B$ is also such that $\braket{B_t}=0$, the above reduces to
\begin{align}
    K_2 = \frac{1}{2i} \int_0^t dt_1 \braket{B_t B_{t_1}} \left[ A_t A_{t_1} - \braket{A_t A_{t_1}}_{1/d} \right] - \text{h.c.} \; .
\end{align}
It corresponds to the Hamiltonian contribution that one would naturally obtain from a TCL expansion at second order~\cite{Colla2024thesis}.

\subsection{Third order}
The expansion at third order, in absence of any further assumption, becomes already more involved. There are four coefficients, which are the following:
\begin{widetext}
\begin{align}
&\Dcals{1}{3} = \Dfds{1}{3} -\Dfds{1}{}\Dfs{2}{3}  -\Dfds{1}{2}\Dfs{3}{}  +\Dfds{1}{}\Dfs{2}{}\Dfs{3}{}\; ,\\
&\Dcal{1}{}{1}{2} = \Dfd{1}{}{1}{2} - \Dfdt{1}{}\Dfs{1}{2} - \Dfds{1}{}\Df{1}{}{2}{} -\Dfd{1}{}{1}{}\Dfs{2}{} -\Dfds{1}{2}\Dft{1}{} \nonumber \\ & \qquad \qquad \qquad + \Dfdt{1}{}\Dfs{1}{}\Dfs{2}{} + 2\Dfds{1}{}\Dfs{2}{}\Dft{1}{} \; ,\\
&\Dcal{1}{2}{1}{} = \Dfd{1}{2}{1}{} - \Dfds{1}{}\Dft{1}{2} - \Dfdt{1}{}\Df{2}{}{1}{} -\Dfd{1}{}{1}{}\Dft{2}{} -\Dfdt{1}{2}\Dfs{1}{} \nonumber \\ & \qquad \qquad \qquad + \Dfds{1}{}\Dft{1}{}\Dfs{2}{} + 2\Dfdt{1}{}\Dft{2}{}\Dfs{1}{}\; , \\
&\Dcalt{1}{3} = \Dfdt{1}{3} -\Dfdt{1}{}\Dft{2}{3}  -\Dfdt{1}{2}\Dft{3}{}  +\Dfdt{1}{}\Dft{2}{}\Dft{3}{} \; .
\end{align}
\end{widetext}
Inserted into \eqref{eq:Kn} give the following third order contribution:
\begin{align}
    K_3 = -\frac{1}{2} \int_0^t dt_1 \int_0^t dt_2 &\Big[ f(t,t_1,t_2)  X(t,t_1,t_2) \\
    & \;  - g(t,t_1,t_2) Y(t,t_1,t_2) + \text{h.c.}\Big] \; ,  \nonumber
\end{align}
where we have defined the coefficients
\begin{align}
    f(t,t_1,t_2) =& \braket{B_t B_{t_1} B_{t_2}}\theta_{t_1^2} - \braket{B_t}\braket{B_{t_1} B_{t_2}}\theta_{t_1^2} \nonumber \\ &- \braket{B_t B_{t_1}}\braket{B_{t_2}} + \braket{B_t}\braket{B_{t_1}}\braket{B_{t_2}} \; ,
    \end{align}
    \begin{align}
    g(t,t_1,t_2) =& \braket{B_{t_1} B_{t} B_{t_2}} - \braket{B_t}\braket{B_{t_1} B_{t_2}} \nonumber \\& - \braket{B_{t_1}B_t}\braket{B_{t_2}}  - \braket{B_t B_{t_2}}\braket{B_{t_1}} \nonumber \\& + 2\braket{B_t}\braket{B_{t_1}}\braket{B_{t_2}} \; ,
\end{align}
and the operators
\begin{align}
    X(t,t_1,t_2) =& A_t A_{t_1} A_{t_2} -\braket{A_t A_{t_1} A_{t_2}}_{1/d} \nonumber \\& - A_{t_1} A_{t_2} \braket{A_t}_{1/d} + A_t \braket{A_{t_1} A_{t_2}}_{1/d} \; ,\\
    Y(t,t_1,t_2) =& A_{t} A_{t_2} \braket{A_{t_1}}_{1/d} - A_{t_1} \braket{A_{t} A_{t_2}}_{1/d} \; . 
\end{align}

Whenever $\Tr{A}=0$ and $\braket{B_t}=0$, the expression simplifies but does not grant further insights. Nonetheless, it is also frequent the case where the mechanism for which $\braket{B_t}=0$ implies that also all odd-ordered cumulants vanish (such as for linearly coupled thermal bosonic environments); in that case, $K_3$ is identically zero -- as is the contribution to the whole generator~\cite{Breuer2007}. 

\subsection{Fourth order}
At fourth order the full expansion becomes quite cumbersome. For this reason, we assume from the start that $\braket{B_t}=0$, which helps reduce the number of terms as we have seen already at lower orders. In particular, first order disappears and all coefficients for $n=2,3$ are simply given by $\Dcal{1}{k}{1}{n-k}= \Dfd{1}{k}{1}{n-k}$. Using the recursive formula \eqref{eq:Dcal-rec} we can then easily compute the fourth order coefficients, which read
\begin{align}\label{eq:D41}
&\Dcals{1}{4} = \Dfds{1}{4} -\Dfds{1}{2}\Dfs{3}{4} \; ,\\
&\Dcal{1}{}{1}{3} = \Dfd{1}{}{1}{3} -\Dfd{1}{}{1}{}\Dfs{2}{3} \nonumber \\ \label{eq:D42}
& \quad \quad \quad \quad \quad \quad-\Dfds{1}{2}\Df{1}{}{3}{} \; , \\
&\Dcal{1}{2}{1}{2} = \Dfd{1}{2}{1}{2} -\Dfd{1}{}{1}{}\Df{2}{}{2}{} 
\nonumber \\ \label{eq:D43}
& \quad \quad \quad \quad \quad \quad-\Dfds{1}{2}\Dft{1}{2} - \Dfdt{1}{2}\Dfs{1}{2} \; , \\
&\Dcal{1}{3}{1}{} = \Dfd{1}{3}{1}{} -\Dfd{1}{}{1}{}\Dft{2}{3}\nonumber \\ \label{eq:D44}
& \quad \quad \quad \quad \quad \quad-\Dfdt{1}{2}\Df{3}{}{1}{} \; \\
&\Dcalt{1}{4} = \Dfdt{1}{4} -\Dfdt{1}{2}\Dft{3}{4}  \,\label{eq:D45}
\end{align}
and give the following contribution to the Hamiltonian:
\begin{align}
    K_4 = -\frac{1}{2i} \int_0^t & dt_1  dt_2 dt_3 \Big[ \bar{f}(t,t_1,t_2,t_3) \bar{X}(t,t_1,t_2,t_3)   \\
    & \; - \bar{g}(t,t_1,t_2,t_3) \bar{Y}(t,t_1,t_2,t_3) - \text{h.c.}\Big] \; . \nonumber
\end{align}
Similarly to what we did for third order, we defined the coefficients
\begin{align}
    \bar{f}(t,t_1,t_2,t_3) =& \braket{B_t B_{t_1} B_{t_2}B_{t_3}}\theta_{t_1^3} \nonumber \\ &- \braket{B_t B_{t_1}}\braket{B_{t_2} B_{t_3}}\theta_{t_2^3} \; ,
\end{align}
\begin{align}
    \bar{g}(t,t_1,t_2,t_3) =& \braket{B_{t_1} B_{t} B_{t_2}B_{t_3}}\theta_{t_2^3} \nonumber \\ &- \braket{B_{t_1}B_t}\braket{B_{t_2} B_{t_3}}\theta_{t_2^3} \nonumber \\ &- \braket{B_t B_{t_2}}\braket{B_{t_1}B_{t_3}} \; ,
\end{align}
and the operators
\begin{align}
    \bar{X}(t,t_1,t_2,t_3) =& A_t A_{t_1} A_{t_2}A_{t_3}-\braket{A_t A_{t_1} A_{t_2}A_{t_3}}_{1/d} \nonumber \\ 
    & - A_{t_1} A_{t_2}A_{t_3} \braket{A_t}_{1/d} \nonumber \\
    &+ A_t \braket{A_{t_1} A_{t_2} A_{t_3}}_{1/d} \; ,\\
    \bar{Y}(t,t_1,t_2,t_3) =&  A_{t} A_{t_2}A_{t_3} \braket{A_{t_1}}_{1/d} \nonumber \\ &- A_{t_1} \braket{A_{t} A_{t_2}A_{t_3}}_{1/d} \; . 
\end{align}
The latter will slightly simplify under the assumption that $\Tr{A}=0$.

The calculation of the coefficients \eqref{eq:D41}-\eqref{eq:D45} was made significantly easier through the use of the recursive formula, in particular because the simplification of lower-order coefficients can be quickly carried through to higher orders. 

\section{Conclusions}

In this work, we have developed a recursive perturbative expansion for the time-convolutionless (TCL) generator of an open quantum system, explicitly formulated in a generalized Lindblad structure using left- and right-acting operators. This approach provides a systematic framework for deriving the generator at arbitrary orders without imposing constraints on the system or environment beyond the assumption of an initially uncorrelated state.

A key advantage of this formulation is its ability to transparently separate the coherent and dissipative contributions of the TCL generator. This decomposition naturally leads to a recursive expression for the effective system Hamiltonian, capturing environment-induced energy renormalization beyond the weak-coupling and Markovian limits. By maintaining a Lindblad-like structure, our method offers insights into non-Markovian and strong-coupling effects, as well as their thermodynamic implications.

To illustrate the effectiveness of this approach, we computed explicit expressions up to fourth order, illustrating how the recursive structure simplifies the calculations by propagating lower-order simplifications to higher orders. Moreover, the tools developed in this work have been used to investigate the structure of coherent dynamics of spin systems at all orders and to corroborate numerical evidence~\cite{Colla2025K}.

This framework paves the way for systematically studying open quantum systems with complex environments while preserving a physically interpretable form of the TCL generator.

\acknowledgments
A.C. acknowledges support from MUR via the PRIN 2022 Project “Quantum Reservoir Computing (QuReCo)” (contract n. 2022FEXLYB). A.C. and H.P.B. acknowledge financial support from the European Union's Framework Programme for Research and Innovation Horizon 2020 (2014-2020) under the Marie Sk\l{}odowska-Curie Grant Agreement No.~847471. G.G. acknowledges financial support by the DAAD, the Deutsche Forschungsgemeinschaft (DFG, German Research Foundation, project numbers 447948357 and 440958198), the SinoGerman Center for Research Promotion (Project M-0294), the German Ministry of Education and Research (Project QuKuK, BMBF Grant No. 16KIS1618K), and the Stiftung Innovation in der Hochschullehre and the Spanish MICIN (project PID2022-141283NB-I00) with the support of FEDER funds.

\begin{appendix}\label{sec:appA}
\section{Details on the generalized cumulants}
The generalized cumulants in eq.~\eqref{eq:Dcal} are obtained by replacing eqns.~\eqref{eq:mu_kn} and \eqref{eq:mu_kn_dot} into \eqref{eq:gen-exp-lambda} and by relabelling the indices on the time variables:
\begin{widetext}
\begin{align}
  \mathcal{D}(\boldsymbol{\tau}_{1}^{k},\boldsymbol{s}_{1}^{n-k})&=\sum_{q=1}^{n}(-)^{q+1}\sum_{\substack{0 \le k_{1}\le ...\le k_{q}= k \\
   0\le m_{1}\le ...\le m_{q}= n-k}}
  \dot{D}(\boldsymbol{\tau}_{1}^{k_{1}},\boldsymbol{s}_{1}^{m_{1}})D(\boldsymbol{\tau}_{1+k_{1}}^{k_{2}},\boldsymbol{s}_{1+m_{1}}^{m_{2}})\dots D(\boldsymbol{\tau}_{1+k_{q}}^{n},\boldsymbol{s}_{1+m_{q}}^{k}) \nonumber\\
  &=\sum_{q=1}^{n}(-)^{q+1}\sum_{\substack{ 0 \le k_{1}\le ...\le k_{q}= k \\
     0\le m_{1}\le ...\le m_{q}= n-k}}
    \dot{D}(\boldsymbol{\tau}_{1}^{k_{1}},\boldsymbol{s}_{1}^{m_{1}})\prod_{j=1}^{q-1} D(\boldsymbol{\tau}_{1+k_{j}}^{k_{j+1}},\boldsymbol{s}_{1+m_{j}}^{m_{j+1}}) \; .
\end{align}
From this, we can again rearrange the terms to explicitly bring out the lower order contributions, in the following way: 
\begin{align}
  \mathcal{D}(\boldsymbol{\tau}_{1}^{k},\boldsymbol{s}_{1}^{n-k}) &=
  \dot{D}(\boldsymbol{\tau}_{1}^{k},\boldsymbol{s}_{1}^{n-k})-
  \sum_{l=0}^{k-1}\sum_{r=0}^{n-k-1}\left(
  \sum_{q=1}^{l+r}(-)^{q+1}\sum_{\substack{ 0 \le k_{1}\le ...\le k_{q}= l \\
     0\le m_{1}\le ...\le m_{q}= r}}
    \dot{D}(\boldsymbol{\tau}_{1}^{k_{1}},\boldsymbol{s}_{1}^{m_{1}})\prod_{j=1}^{q-1} D(\boldsymbol{\tau}_{1+k_{j}}^{k_{j+1}},\boldsymbol{s}_{1+m_{j}}^{m_{j+1}})\right)
    D(\boldsymbol{\tau}_{l}^{k},\boldsymbol{s}_{r}^{n-k}) \; .
\end{align}

which gives the recursive formula eq.~\eqref{eq:Dcal-rec}.
\end{widetext}

\end{appendix}

\bibliography{biblio}

\end{document}